\documentclass[%
 aip,
 amsmath,amssymb,
 reprint,%
]{revtex4-1}

\usepackage{graphicx}
\usepackage{dcolumn}
\usepackage{bm}

\usepackage[utf8]{inputenc}
\usepackage[T1]{fontenc}
\usepackage{mathptmx}
\usepackage{etoolbox,color}

\makeatletter
\def\@email#1#2{%
 \endgroup
 \patchcmd{\titleblock@produce}
  {\frontmatter@RRAPformat}
  {\frontmatter@RRAPformat{\produce@RRAP{*#1\href{mailto:#2}{#2}}}\frontmatter@RRAPformat}
  {}{}
}%
\makeatother
\begin{document}

\preprint{AIP/123-QED}

\title{\textbf{Effect of cross-sectional anisotropy on shock train dynamics in supersonic internal flows} 
}%

\author{J. Singh}
 \email{Contact author: jagmohan@umich.edu}
 \affiliation{Department of Aerospace Engineering, University of Michigan, Ann Arbor}
\author{V. Raman}%
\affiliation{Department of Aerospace Engineering, University of Michigan, Ann Arbor}%

\date{\today}

\begin{abstract}
 
 This study investigates the effect of duct aspect ratio ($AR$), defined as the ratio of major to minor axis in an elliptical duct, on shock train dynamics for a freestream Mach number of 2.1. The $AR$ is varied from $1.0$ to $3.0$ while maintaining a constant cross-sectional area and identical upstream conditions, thereby ensuring the same inlet mass flow and nearly constant boundary-layer-induced blockage across all $AR$. This isolates shape-induced confinement effects. Simulations are performed using an embedded-boundary method with adaptive mesh refinement which enabled a finest resolution of 48$\mu$m resolving the shocks in the shock train. The results show that increasing $AR$ significantly modifies the shock train morphology. The number of discrete shock cells decreases, and the leading shock front elongates along the major axis while contracting along the minor axis. The normal shock stem prominent in the circular duct ($AR=1.0$) nearly disappears at $AR=3.0$. Despite these morphological changes, the wall-pressure trace  and stagnation-pressure loss remain largely insensitive to $AR$. These results indicate that while duct cross-section governs the detailed shock train structure, the overall efficiency of flow deceleration and pressure recovery is dictated primarily by the effective blockage imposed by the turbulent boundary layer, rather than the aspect ratio itself for a given mass flow rate and pressure ratio across the pseudo shock.
 \end{abstract}

\maketitle


\section{\label{sec:introduction}Introduction}

Shock trains are flow structures encountered in confined supersonic internal flows where a high-speed supersonic flow must decelerate and adapt to downstream subsonic conditions \cite{matsuo1999shock,carroll1990characteristics}. A shock train consists of a series of oblique and normal shocks interacting with the turbulent boundary layer along the duct walls, often accompanied by complex expansions and boundary layer separation due to the shocks in the shock train \cite{matsuo1999shock,carroll1990characteristics,hunt2018shock}. The interplay between shock waves and boundary layer results in a highly three-dimensional, unsteady flow field known as pseudo shock when combined with the subsequent mixing region \cite{matsuo1999shock}. Although majority of pressure rise occurs through the shock train, some additional pressure recovery occurs in this subsonic mixing region. Understanding the behavior and characteristics of shock trains, including their length, unsteadiness, and stagnation pressure loss, are critical for the design and operation of various high-speed internal flow systems where reliable and efficient flow deceleration and pressure adaptation are critical. 

Previous studies have extensively investigated the effect of various flow parameters on the shock train behavior. It is known that for Mach numbers $M \lesssim  1.7$, the shock train is dominated by a bifurcating normal shock front~\cite{matsuo1999shock}. As the Mach number increases, the central normal shock stem shrinks, and by $M \gtrsim 2.2$, the system transitions to an oblique shock train, in which inclined shocks distribute the pressure more gradually \cite{hunt2018shock}.  Back pressure also affects the shock train characteristics. Higher back pressures promote normal shock train, while lower back pressures favor oblique shock train \cite{hunt2018shock}. Moreover, as back pressure increases, the total number of shock cells decreases, although the spacing between the first few shocks in the shock train remains largely constant \cite{singh2024shock}. When the back pressure becomes sufficiently high, the shock system can propagate upstream and be expelled from the inlet. This results in a complete breakdown of the system, as the formation of a strong bow shock near the inlet severely restricts the incoming mass flow. Freestream turbulence intensity generally has minimal effect on shock train structure~\cite{cox2001effect}.

In addition to Mach number and back pressure, boundary layer and geometric confinements have been found to significantly affect the shock train dynamics with thicker boundary layer leading to longer and more complex shock train with increased three-dimensionality \cite{fievet2017numerical, singh2024shock, hunt2018shock, waltrup1973structure}. Investigations of square ducts and channels have shown that sidewalls strongly affect the shock train dynamics, with the associated blockage leading to a substantial increase in shock train length \cite{gillespie2023numerical}. This effect is consistent with increasing the boundary layer thickness for a given geometry \cite{vane2013simulations,singh2024shock}. Compared to axisymmetric configurations, rectangular ducts present additional complexities due to low-momentum corner flows and secondary vortices, which contribute to shock train unsteadiness and greater deviations from quasi-one-dimensional predictions \cite{carroll1990characteristics,morgan2014large}. Attempts have been made to study duct aspect-ratio effects in rectangular geometries, however, in those cases, the aspect ratio was varied by modifying a single cross-sectional dimension, which also altered the blockage ratio defined as the ratio of the area occupied by the boundary layer to the cross-sectional area, as well as the incoming mass flow rate making it difficult to isolate the effect of aspect ratio alone \cite{geerts2016shock,cox2001effect}. Significant efforts have been devoted to understand the origin of communication pathways in shock train systems within both rectangular and circular ducts \cite{leonard2021investigation,hunt2019origin}.

Elliptical ducts remain relatively underexplored compared to circular and rectangular configurations.  Unlike rectangular ducts, they lack sharp corners that typically promote corner separation and associated secondary flows, yet they also depart from the azimuthal symmetry of circular ducts, introducing geometric anisotropy that can influence flow development. For a fixed global blockage ratio defined as the fraction of the cross-sectional area occupied by the boundary layer, elliptical geometries inherently support unequal boundary-layer thicknesses relative to the duct axes. These variations lead to differences in local confinement along the azimuthal direction. This feature of elliptical geometries is leveraged in the current study to study whether the pressure recovery in a shock train is governed primarily by the local blockage along the duct’s minor axis, where confinement is most significant, or by the overall blockage integrated over the cross section. 

To investigate this, numerical simulations are conducted using an adaptive mesh refinement (AMR) which allows accurate resolution of shock-boundary layer interactions while ensuring computational efficiency. The duct aspect ratio ($AR$), defined as the ratio of major to minor axis lengths, is varied from 1.0 to 3.0, while maintaining a constant cross-sectional area and uniform inlet boundary layer thickness across all cases. This approach ensures that the mass flow rate and global blockage remain consistent, isolating the role of cross-sectional anisotropy. Simulations are performed at a freestream Mach number of 2.1. The analysis focuses on the shock train structure, turbulent fluctuations, pressure increase and stagnation pressure loss across the shock train. 

\section{Method}
\label{sec.method}
The following governing equations for mass, momentum, energy and species transport are solved in the Cartesian frame of reference:
\begin{align}
\label{eq:gov}
\frac{\partial \rho}{\partial t} + \frac{\partial (\rho u_i)}{\partial x_i} & =   0 \nonumber \\
\frac{\partial (\rho u_i)}{\partial t}  + \frac{\partial (\rho u_i u_j) }{\partial x_i}  = & - \frac{\partial p }{\partial x_j}  +  \frac{\partial \tau_{ij} }{\partial x_j}  \nonumber \\
\frac{\partial (\rho E) }{\partial t}  + \frac{\partial (\rho u_i E+ pu_i)}{\partial x_i} & = \frac{\partial}{\partial x_j} \bigg (\alpha \frac{\partial T}{ \partial x_j} \bigg) + \frac{\partial (\tau_{ij} u_i)}{\partial x_j}  \\ \nonumber
\frac{\partial (\rho Y_k) }{\partial t}  + \frac{\partial (\rho u_i Y_k)}{\partial x_i} & = \frac{\partial}{\partial x_j} \bigg (\rho D \frac{\partial Y_k}{ \partial x_j} \bigg)   \quad \textrm{with} \quad k \in  [1, N_s]
\end{align}
In Eq.~(\ref{eq:gov}), $t$ is time, $u_i$ is the fluid velocity, $x_i$ is the spatial coordinate, $\rho$ is the fluid density, $\alpha$ is fluid's thermal conductivity, $\tau_{ij} = \frac{-2}{3} \mu \frac{\partial u_k}{\partial x_k} \delta_{ij} + \mu \left( \frac{\partial u_j}{ \partial x_i} +  \frac{\partial u_i}{\partial x_j} \right)$ is the viscous stress tensor, $\delta_{ik}$ is the Kronecker delta ($\delta_{ij}=1$ when $i=j$, 0 otherwise), $E = E(x, y, z, t)$ is the total energy of the system, $Y_k$ is the mass fraction of $k^{\textrm{th}}$ species and $N_s$ is the total number of species in the chemical mechanism. Here a two-species mechanism for the working medium air is used. The governing equations are discretized using the finite volume method and are solved using an in-house code \cite{sharma2024amrex} that uses AMReX library \citep{AMReX_JOSS} for mesh handling, parallelization over CPUs and GPUs, and adaptive mesh refinement. Euler flux is computed following the Harten-Lax-van Leer-Contact (HLLC) scheme \cite{toro1994restoration} whereas the diffusive terms are calculated using a second-order central scheme. Temporal integration uses the strong stability preserving second-order Runge-Kutta scheme \cite{gottlieb1998total}. The specific heat, enthalpy of a species, and specific heat ratio of the air are calculated via NASA polynomials.  More details about the solver numerics and its verification can be found in \citet{sharma2024amrex}.

The computational domain consists of an elliptical duct with aspect ratio $AR = a/b$, where $a$ and $b$ are the semi-major and semi-minor axes of the elliptical cross-section, respectively. The aspect ratio $AR$ is varied while maintaining a constant cross-sectional area to ensure that the inlet mass flow rate remains the same across all cases. However, this modifies the duct surface area $\mathcal{S}$ as shown in Table~\ref{tab.cellCount}, consequently the hydraulic diameter $D_h=4V/\mathcal{S}$ or $D_h=4A/\mathcal{P}$ where $V$ is the fluid volume, $\mathcal{S}$ is the surface area, $A$ is the cross-sectional area and $\mathcal{P}$ is the duct perimeter. The flow direction is along the $x-$axis, with the duct’s major axis aligned with the $y-$axis and the minor axis with the $z-$axis. The computational domain extents in the $y-$ and $z-$directions are adjusted according to $AR$ to fully contain the duct's cross-section. The computational domain extends approximately $33R$ in the streamwise direction, with the shock train positioned around $13R$ from the inlet where $R$ is the radius of circular duct ($AR = 1.0$), once it reaches a quasi steady state. The radius of the circular duct ($AR=1.0$) is 1.5~cm. In the following analysis, a shifted streamwise coordinate $x^*$ is used such that the origin $x^* = 0$ corresponds to the location of the shock train, with $x^*$ normalized by $R$. The location of the shock train is determined by identifying the minimum $x-$coordinate along the duct centerline where the local mean streamwise pressure increases rapidly. The non-dimensional time $t^*$ is expressed as $t^*=t U_\infty/R$ where $U_\infty$ is the freestream velocity upstream of the shock train.

\begin{figure}[!h]
	\centering
	\begin{tabular}{c}
         \makebox[0pt]{\raisebox{26ex}{\hspace{-5.8ex}{(a)}}} \includegraphics[trim={8cm 0.0cm 5cm 0},clip,width=.35\textwidth]{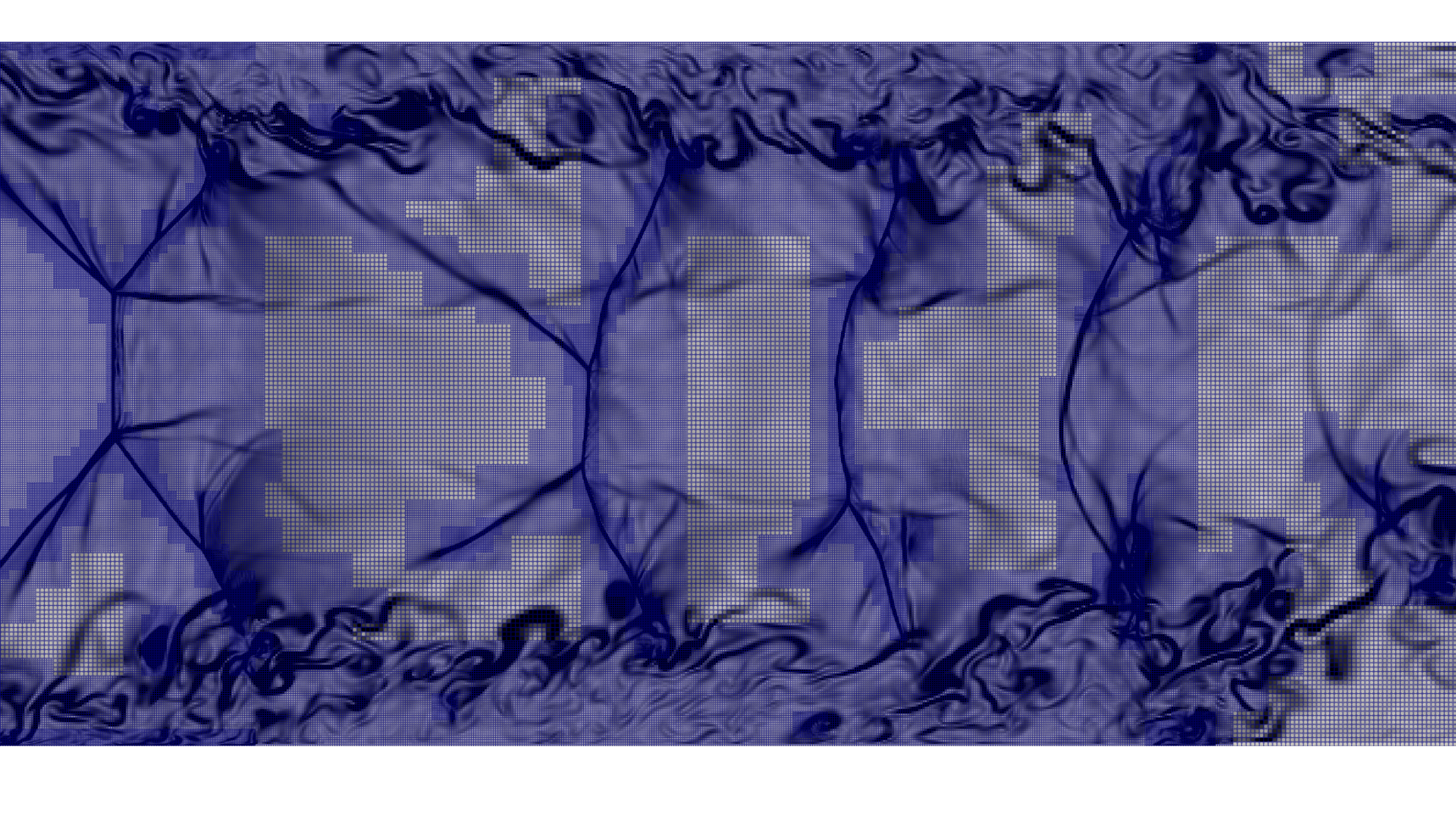} \\ \hspace{10mm} 
        \makebox[0pt]{\raisebox{26ex}{\hspace{-5.8ex}{(b)}}}  \includegraphics[trim={25cm 25.0 20cm 460},clip,width=.25\textwidth]{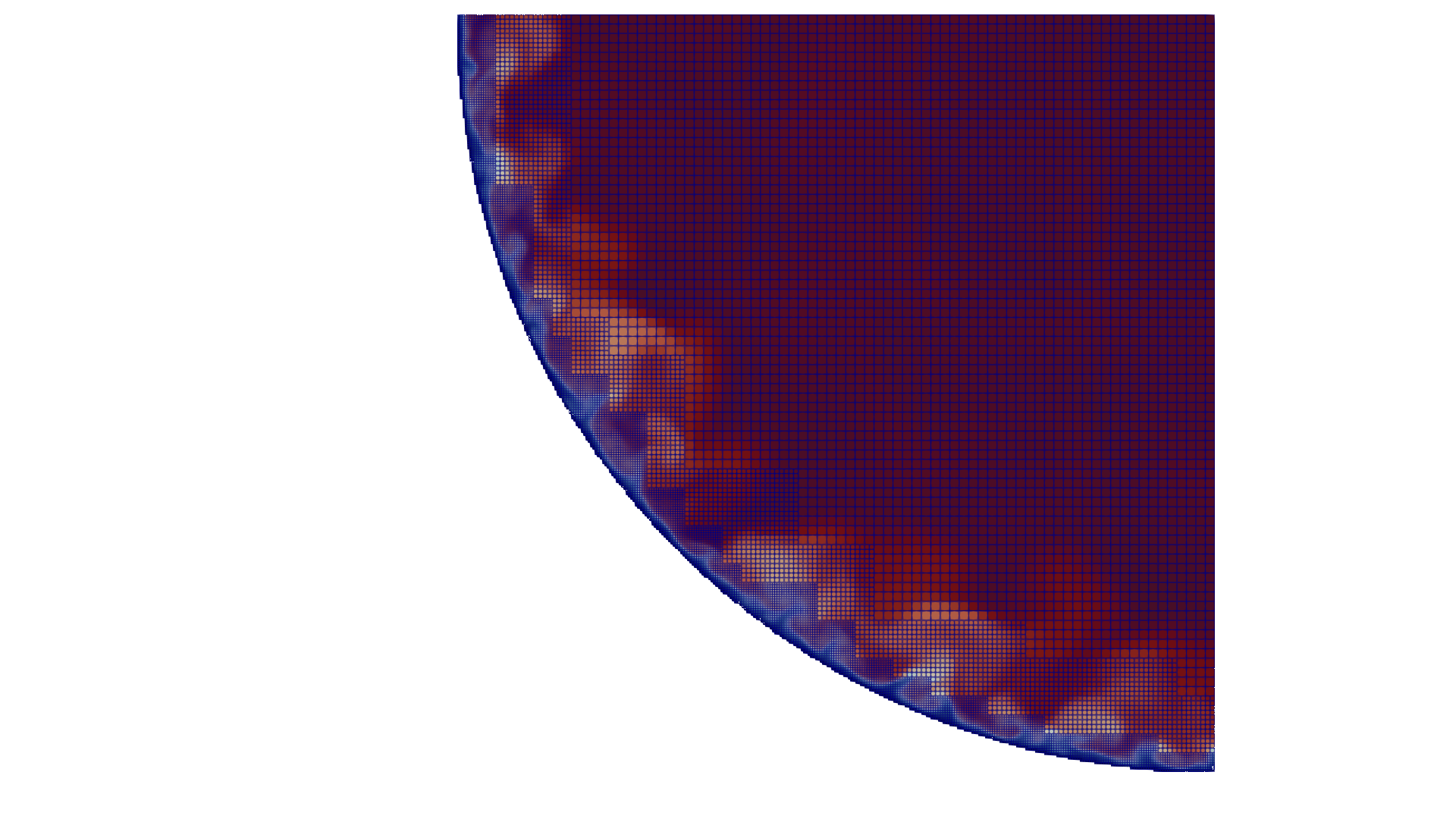} 
        \end{tabular}
	\caption{Refinement via AMR shown on (\textit{a}) an axial plane and (b) a section of the x-plane at $x^* \approx 2.5$ i.e. approximately $2.5R$ upstream of the shock train plotted for $AR=1$. Three levels of refinement can be seen while the coarsest level is not visible in this view.}
	\label{fig.mesh}
\end{figure}

The back pressure is specified by a conical plug positioned at the duct exit similar to the experimental setup of Hunt et al.~\cite{hunt2018shock,hunt2019origin}. For the $AR = 1.0$ case, the duct radius $R$ is 1.5~cm. The conical plug introduces a 16\% blockage, corresponding to a throat-to-cross-sectional area ratio of 84\%. The cone dimensions are held fixed for all aspect ratios. As the plug geometry is resolved on the level 0 mesh, the effective blockage differs slightly compared to the exact value.  According to isentropic flow relations, if the ratio of throat area to cross-sectional area $A_{\text{throat}}/A$ is kept constant across all aspect ratios, the same back pressure is expected for all AR cases assuming ideal isentropic conditions and neglecting viscous effects and shock-induced losses between the downstream end of the shock train and the throat. 

Close to the outlet, an expansion zone is added to ensure that the flow is fully supersonic at the outlet where first-order extrapolation boundary conditions are used for all variables. Duct walls and the plug are represented using the EB method \cite{schwartz2006cartesian}, which classifies computational cells into regular cells, covered cells, and cut cells. Regular cells lie entirely within the fluid domain and are treated with standard finite-volume discretization. Covered cells are fully contained within the solid geometry, and no computations are performed on them. Cut cells are partially occupied by the embedded boundary. For the cut cells, numerical fluxes are modified to enforce boundary conditions at the fluid–solid interface. The duct walls and the conical plug are modeled as adiabatic, no-slip, non-penetrating boundaries, while the boundaries in the expansion zone of the computational domain are treated as slip, non-penetrating walls. The inlet boundary layer thickness $\delta_{in}$ is set to 5\% of the radius of the circular duct $R$ corresponding to $AR = 1.0$. Air is used as a working medium with an inlet Mach number of 2.1 and a temperature of 700K. As the boundary layer develops, its thickness increases to approximately $0.11R$ at a location roughly one shock train length upstream of the leading shock. At this position, the momentum thickness is  $\theta \approx 0.015R$, corresponding to a momentum-thickness-based Reynolds number of $Re_\theta = \rho_w U_\infty \theta /\mu_w \approx 626$.

\begin{table}[!ht]
\centering
\begin{tabular}{ccccccc}
$AR$ & $\mathcal{S}_{AR}/\mathcal{S}_{AR=1.0}$ & $\textrm{N}_{\textrm{Level-0}}$ & $\textrm{N}_{\textrm{Level-1}}$ & $\textrm{N}_{\textrm{Level-2}}$ & $\textrm{N}_{\textrm{Level-3}}$ & $\textrm{N}_{\textrm{Total}}$ \\
\hline
1.0  & 1.0 &13 & 59 & 145 & 297 & 514\\
2.0 &  1.08 & 11 & 62 & 139 & 327 & 539 \\
3.0 &  1.16 &9 & 53 & 124 & 305 & 491 \\
\end{tabular}
\caption{Duct surface area, $\mathcal{S}$, normalized by the surface area of $AR=1.0$ and final cell count $\textrm{N}$ (in millions) at different mesh refinement levels and the total cell count for different aspect ratios $AR$.}
\label{tab.cellCount}
\end{table}

The computational domain is discretized using a structured and uniform grid with four levels of mesh refinement. Mesh is refined dynamically  based on local pressure gradients to capture shock features and based on cut cell criteria to resolve the walls and boundary layers. The finest grid level has a grid spacing of 48 $\mu$m while the coarsest grid level has 375 $\mu$m. Each successive mesh level is refined by a factor of two in all spatial directions. To manage computational cost, the mesh is coarsened downstream of the shock train where fine resolution is no longer required. Table~\ref{tab.cellCount} summarizes the number of cells at each mesh level at the end of the simulations, and Fig.~\ref{fig.mesh} shows the mesh refinement in regions containing shocks. The total cell count in each simulation is approximately 600 million (see Table~\ref{tab.cellCount}). The simulations are run on NVIDIA H100 GPUs, with the number of GPUs dynamically increased to accommodate the growing cell count introduced by AMR as the simulation progresses. Once a quasi–steady state is reached, the simulations used 15 GPUs, sustaining a workload of ~35 million cells per GPU, with each coarse time step advancing in approximately 27 seconds.   
  
\section{Results}

\subsection{Upstream flow}
\begin{figure}[!h]
	\centering
		  \includegraphics[trim={0cm 0cm 0cm 0},clip,width=.49\textwidth]{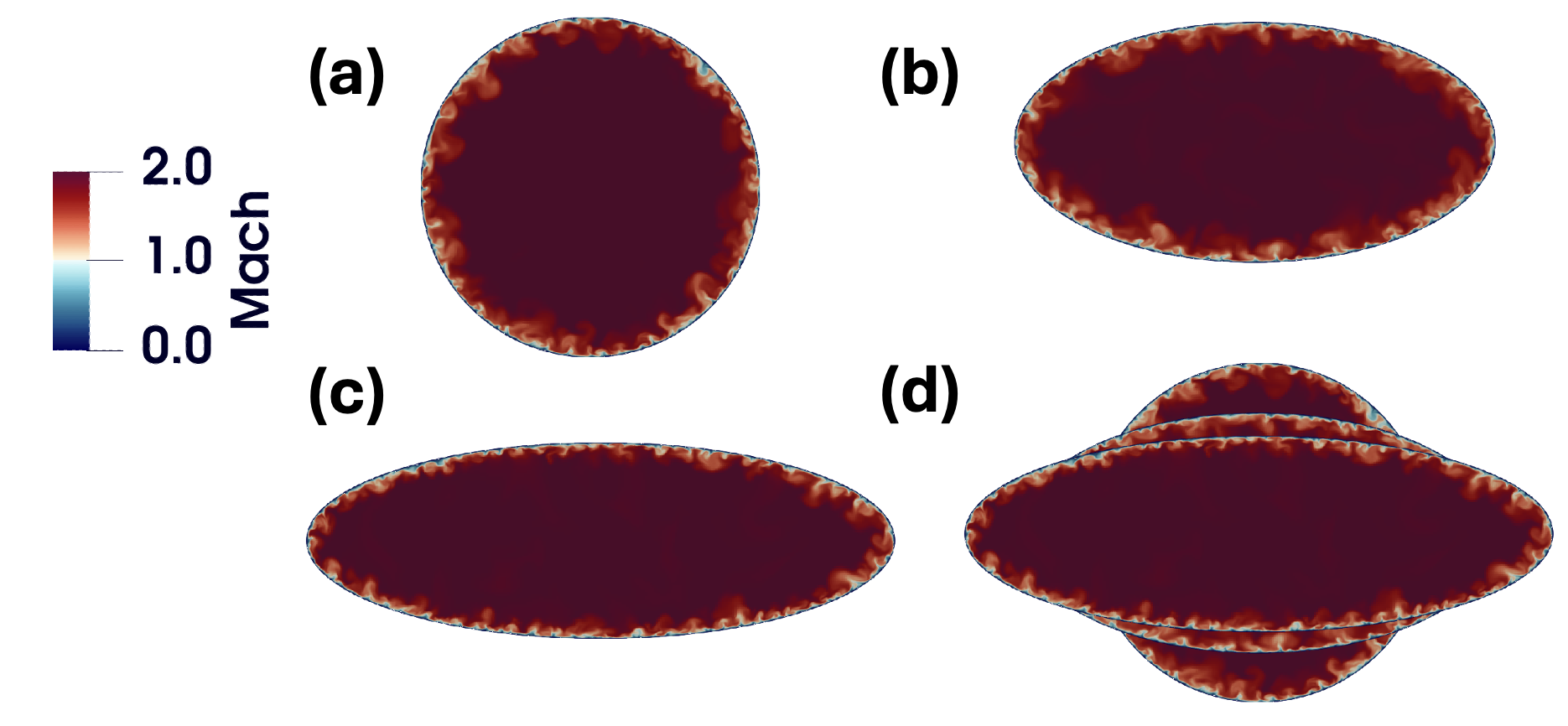} 
	\caption{Instantaneous Mach number upstream of the shock train plotted at $x^*\approx 1.6$ or at a distance of 0.25 times the shock train length for (\textit{a}) $AR=1.0$ (\textit{b}) $AR=2.0$ and (\textit{c}) $AR=3.0$. Panel (\textit{d}) overlays the cross-sections for all aspect ratios to highlight the differences in the cross-sections relative to each other.}
	\label{fig.instant_xupstream}
\end{figure}

\begin{figure}[!ht]
	\centering
	\begin{tabular}{c}
		\makebox[0pt]{\raisebox{32ex}{\hspace{-2.4ex}{(a)}}}  \includegraphics[trim={0cm 0.1cm 0cm 0},clip,width=.44\textwidth]{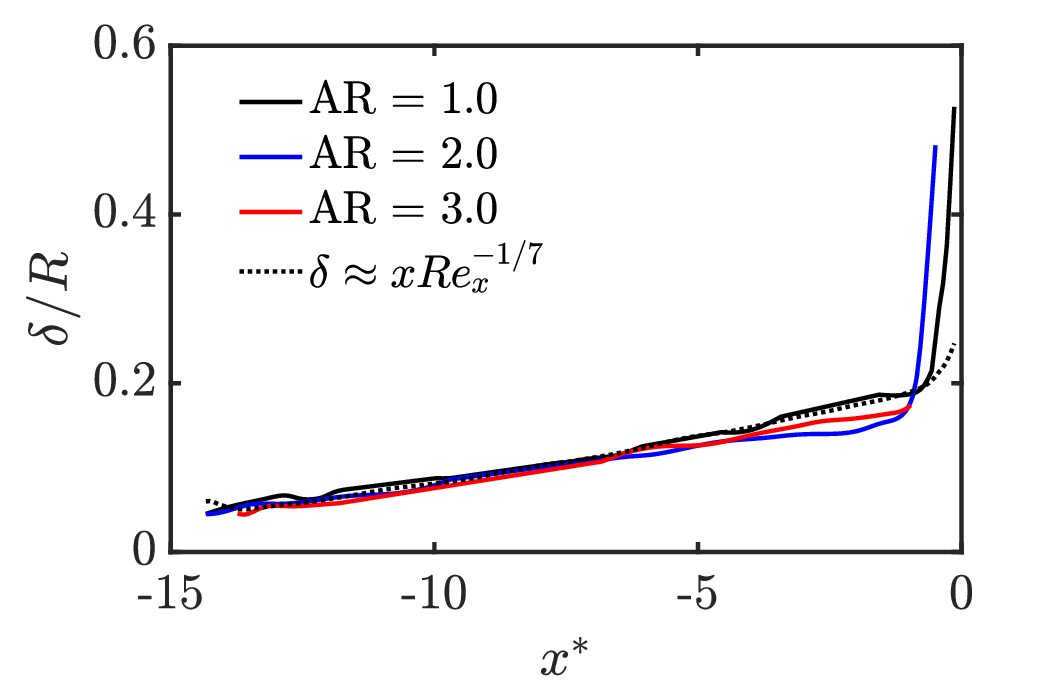}  \\ \hspace{2mm}
        \makebox[0pt]{\raisebox{32ex}{\hspace{-2.4ex}{(b)}}}  \includegraphics[trim={0cm 0.0cm 0cm 0},clip,width=.44\textwidth]{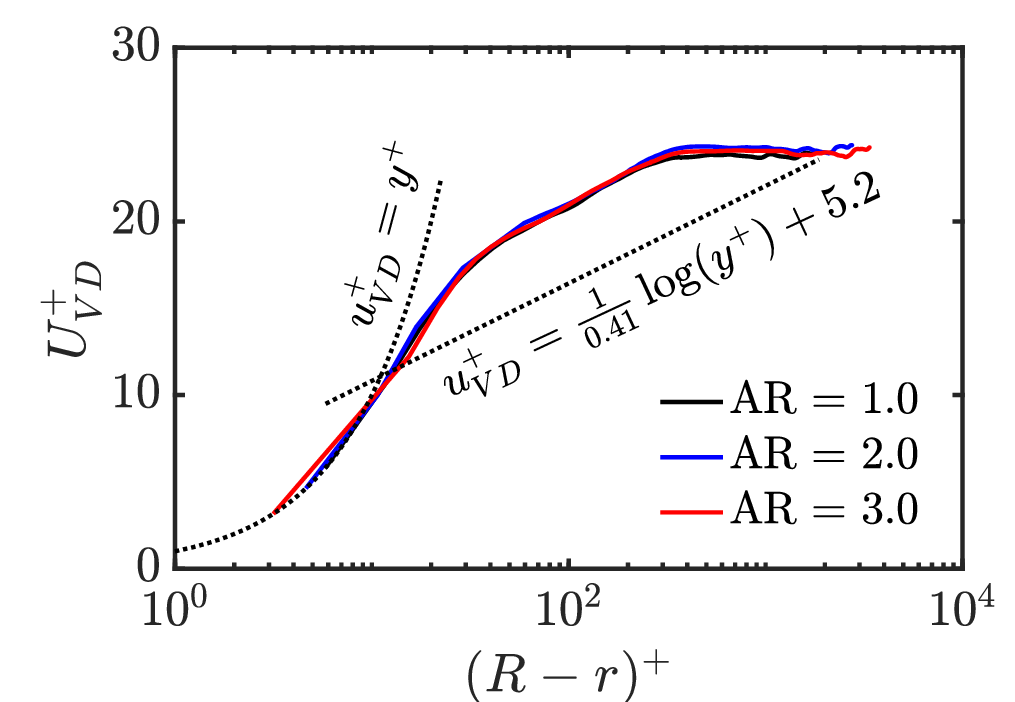}
        \end{tabular}
	\caption{ (\textit{a}) Boundary layer thickness upstream of the shock train. (\textit{b}) Mean velocity profiles plotted in wall-units for different aspect ratios at $2.6R$ upstream of the shock train. }
	\label{fig.bl_vel}
\end{figure}

The instantaneous Mach number contours in Fig.~\ref{fig.instant_xupstream} show fine-scale near-wall structures upstream of the shock train indicating well-developed turbulent boundary layer. Figure~\ref{fig.instant_xupstream}(\textit{d}) overlays the cross-section for all $AR$ to highlight the differences in the cross-sections of different $AR$ relative to each other. The similarity of these features across all aspect ratios suggests that the upstream boundary layer exhibits similar turbulent behavior across the cases considered. Figure~\ref{fig.bl_vel}(a) quantifies this observation by showing the streamwise evolution of the boundary layer thickness, defined as the wall-normal distance at which the mean streamwise velocity $U_x$ reaches 99\% of the freestream value. Across all aspect ratios, the boundary layer growth follows the classical Prandtl power-law scaling $\delta \sim x Re_x^{-1/7}$, consistent with fully developed turbulent boundary layer behavior.

This similarity is further confirmed by the mean velocity profiles presented in Fig.~\ref{fig.bl_vel}(b), plotted in wall units using the Van Driest transformation:
\begin{equation}
    u_{VD}^+ = \int_0^{u^+} \sqrt{\rho/\rho_w}\, du^+.
    \label{eq.vd}
\end{equation}
In Eq.~\ref{eq.vd} the velocity is normalized by the friction velocity $u_\tau = \sqrt{\tau_w / \rho_w}$, with the wall shear stress is defined as $\tau_w = \mu_w (dU/dy)|_w$, and $\rho_w$ and $\mu_w$ represent the time-averaged density and viscosity at the wall. The transformed velocity profiles for different $AR$ collapse onto a single curve, indicating that the near-wall turbulence structure remains comparable across the cases considered. The profiles follow the expected slope in the logarithmic region, although the intercept is higher than the classical value reported for incompressible turbulent boundary layers. This overestimation in Van-Driest transformed coordinates is a well-documented characteristic of supersonic compressible flows \cite{modesti2019direct, ghosh2010compressible, babbar2025supersonic}, arising from the additional influence of compressibility effects in axisymmetric ducts.

\subsection{Shock train dynamics}
\begin{figure*}[!ht]
	\centering
	\begin{tabular}{cc}
		\makebox[0pt]{\raisebox{16ex}{\hspace{-2.8ex}{(a)}}}  \includegraphics[trim={0cm 0.5cm 0cm 0},clip,width=.42\textwidth]{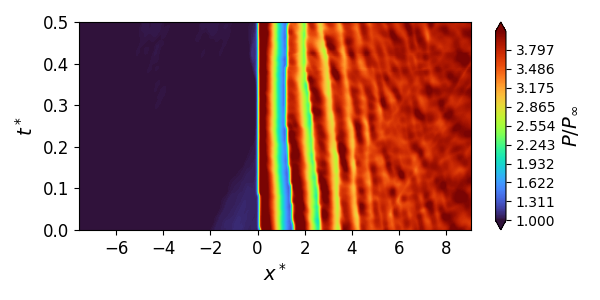}  & \hspace{2mm}
        \makebox[0pt]{\raisebox{16ex}{\hspace{-2.8ex}{(b)}}}  \includegraphics[trim={0cm 0.5cm 0cm 0},clip,width=.42\textwidth]{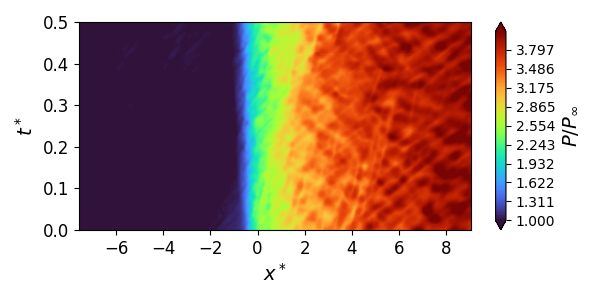} \\
        \makebox[0pt]{\raisebox{16ex}{\hspace{-2.8ex}{(c)}}}  \includegraphics[trim={0cm 0.5cm 0cm 0},clip,width=.42\textwidth]{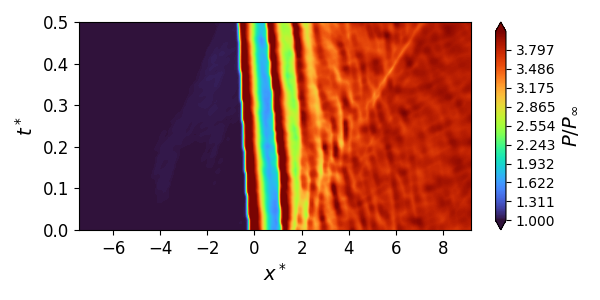}  & \hspace{2mm}
        \makebox[0pt]{\raisebox{16ex}{\hspace{-2.8ex}{(d)}}}  \includegraphics[trim={0cm 0.5cm 0cm 0},clip,width=.42\textwidth]{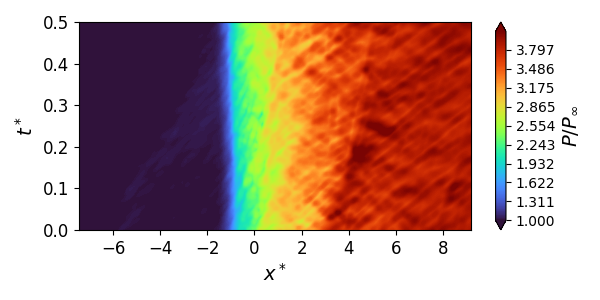} \\
        \makebox[0pt]{\raisebox{16ex}{\hspace{-2.8ex}{(e)}}}  \includegraphics[trim={0cm 0.5cm 0cm 0},clip,width=.42\textwidth]{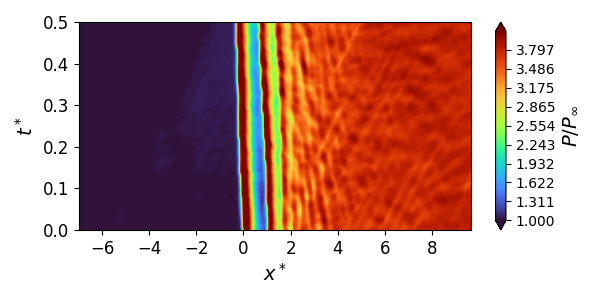}  & \hspace{2mm}
        \makebox[0pt]{\raisebox{16ex}{\hspace{-2.8ex}{(f)}}}  \includegraphics[trim={0cm 0.5cm 0cm 0},clip,width=.42\textwidth]{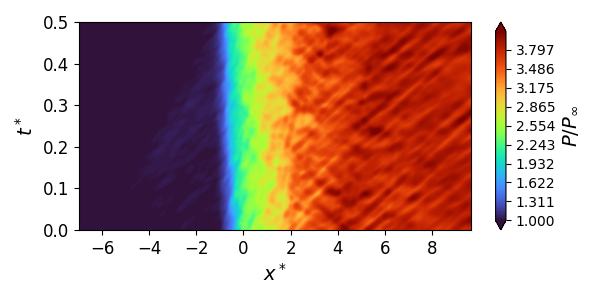}      
        \end{tabular}
	\caption{ History of the pressure at the (a,c,e) centerline and (b,d,f) at the wall plotted for (a,b) $AR=1.0$ (c,d) $AR=2.0$ and (e,f) $AR=3.0$.}
	\label{fig.history}
\end{figure*}

Figure~\ref{fig.history} shows the time‑history of the normalized instantaneous static pressure, $P/P_{\infty}$ where $p_\infty$ is the freestream pressure upstream of the shock train, along the duct centerline and at the wall for different aspect ratios. The centerline pressure contours (Fig.~\ref{fig.history}a,c,e) show a sequence of alternating high‑ and low‑pressure bands downstream of the shock‑train origin $(x^{*}=0)$, which are characteristic of the discrete shocks that form the internal structure of the shock train. The spacing between these shock fronts remains approximately uniform over time which indicates quasi‑steady behavior in which the internal structure of the shock train remains statistically steady even as its absolute position shifts, The entire shock system drifts upstream gradually with the drift speed increasing slightly with aspect ratio ($\approx0.015U_{\infty}$ for AR=1.0 vs. $\approx 0.021 U_{\infty}$ for AR=3.0). Similar upstream drift of the shock train has been observed in other numerical  studies using very fine resolution~\cite{fievet2017numerical,gillespie2023numerical}. Spatiotemporal oscillations appear as inclined wavefronts in the $x^{*}-t^{*}$ plane downstream of $x^{*}\approx0$ at the axis (Fig.~\ref{fig.history}a,c,e) and at the wall (Fig.~\ref{fig.history}b,d,f), showing both upstream and downstream propagation of pressure disturbances. The upstream‑propagating disturbances are linked with shock–boundary layer feedback mechanisms that sustain the shock train oscillations and influence its length and dynamics ~\cite{fievet2017numerical,hunt2019origin,leonard2021investigation}. 

\begin{figure*}[!ht]
	\centering
	\begin{tabular}{ccc}
		\makebox[0pt]{\raisebox{10ex}{\hspace{-2.8ex}{(a)}}}  \includegraphics[trim={26cm 14cm 27cm 13cm},clip,width=.23\textwidth]{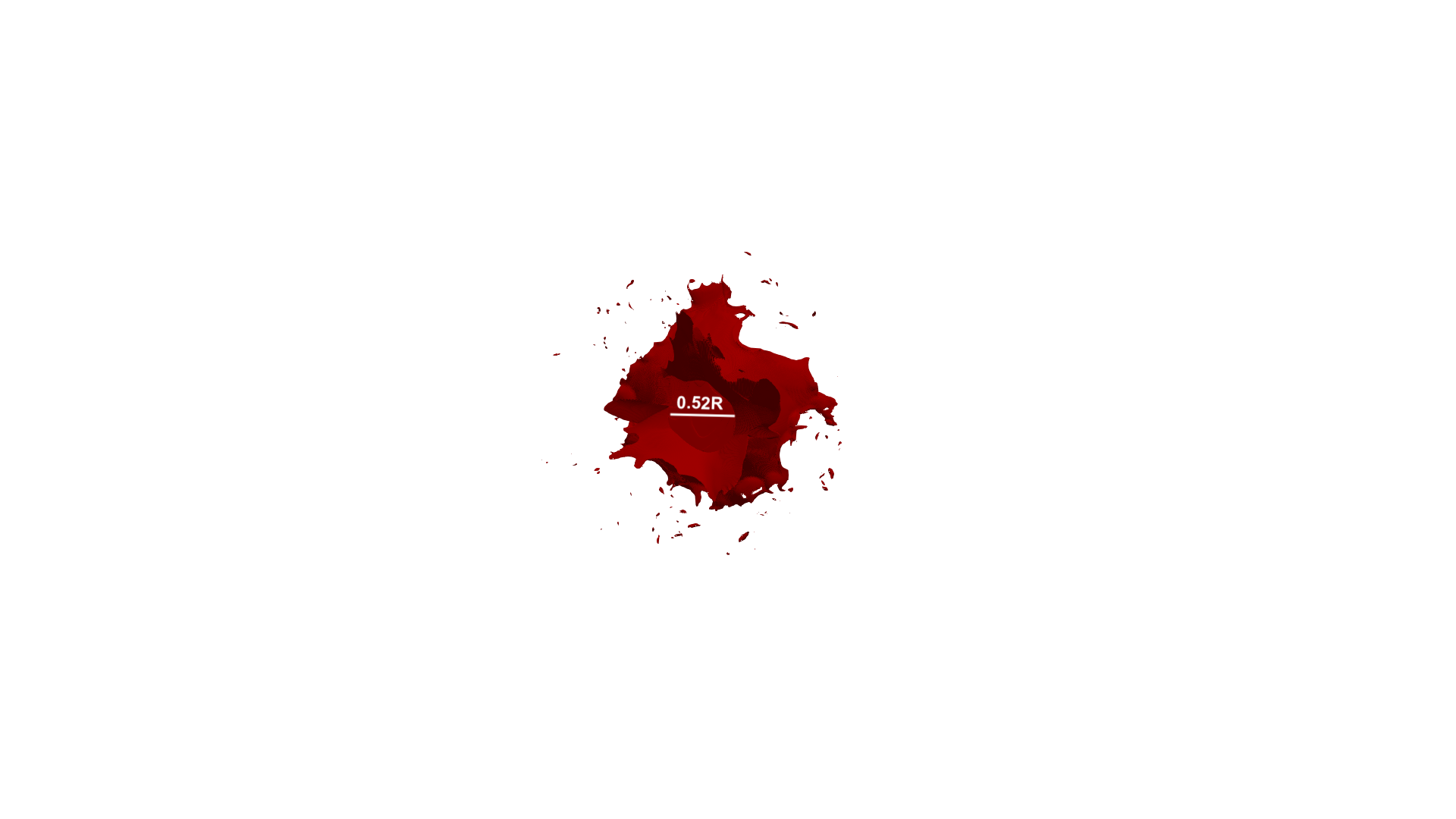}  & \hspace{10mm}
        \makebox[0pt]{\raisebox{10ex}{\hspace{-2.8ex}{(b)}}}  \includegraphics[trim={22cm 14cm 25cm 13cm},clip,width=.31\textwidth]{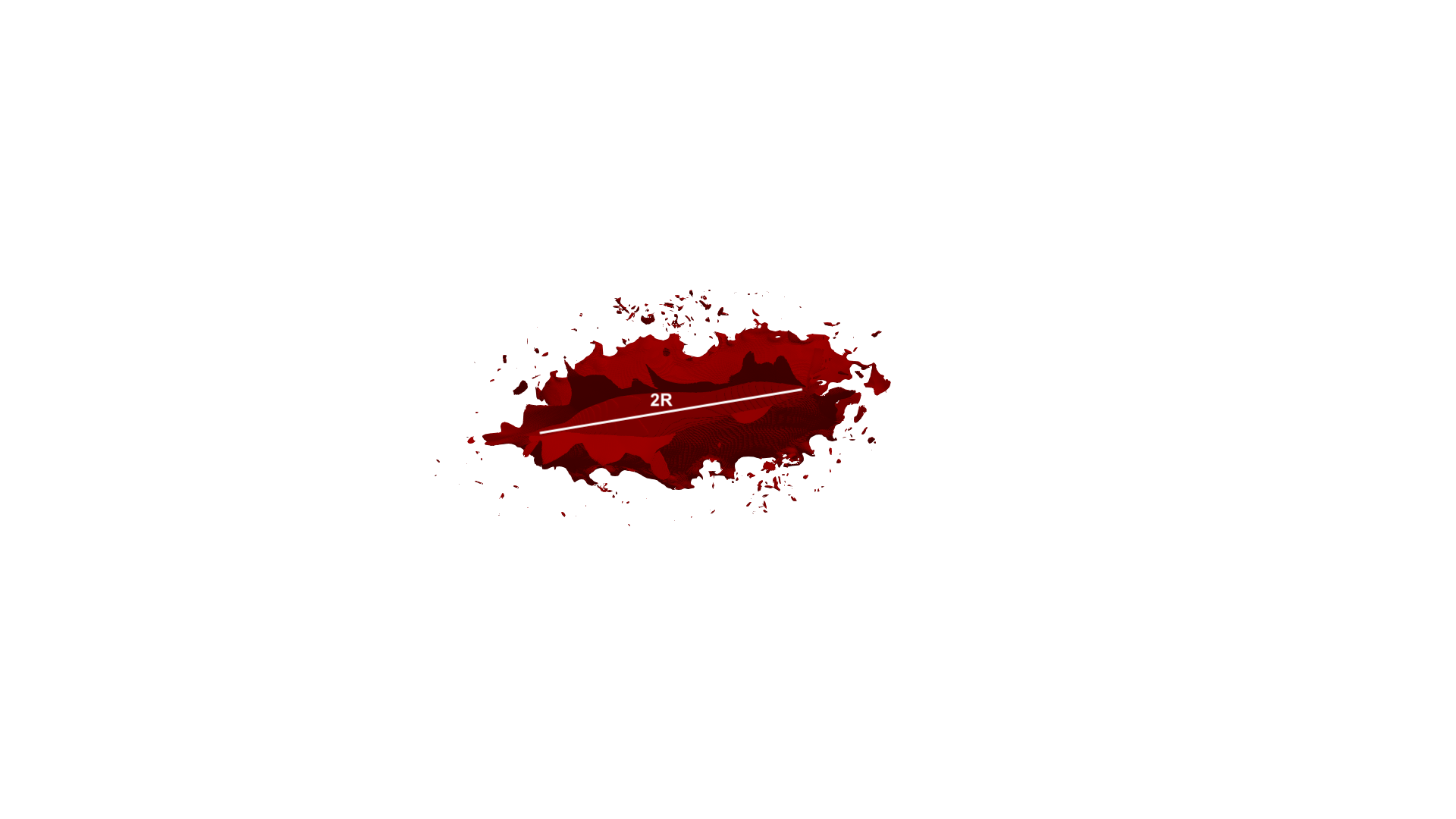}  & \hspace{10mm} 
        \makebox[0pt]{\raisebox{10ex}{\hspace{-2.8ex}{(c)}}}  \includegraphics[trim={18cm 14cm 20cm 14cm},clip,width=.42\textwidth]{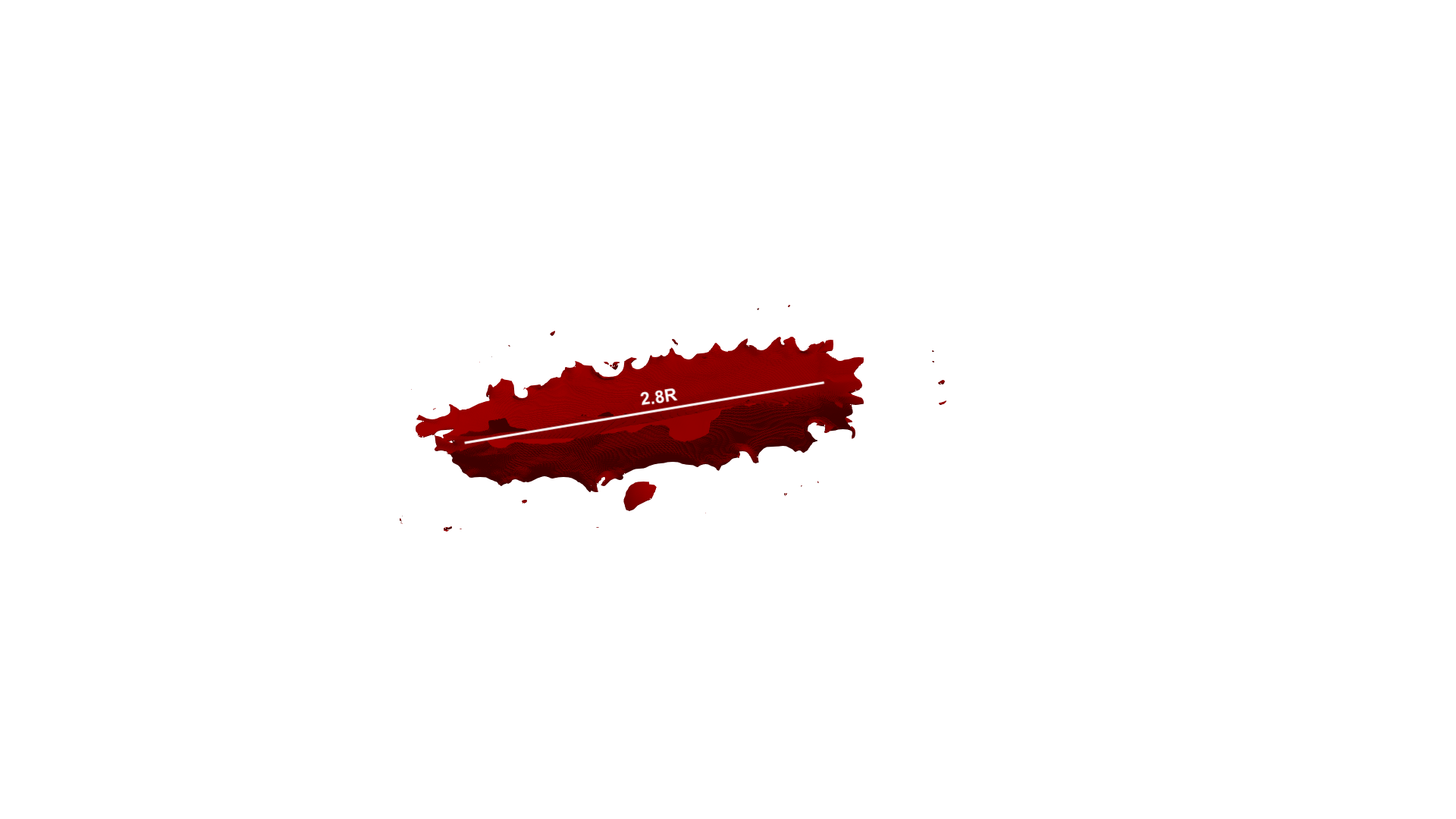}
        \end{tabular}
	\caption{ Instantaneous three-dimensional structure of the leading shock for (a) $AR=1.0$ (b) $AR=2.0$ and (c) $AR=3.0$.}
	\label{fig.instant3D}
\end{figure*}

The effect of geometric confinement on the spatial development of the leading shock is shown in Fig.~\ref{fig.instant3D}, which presents the instantaneous three-dimensional structure of the leading shock for different aspect ratios, visualized using iso-surfaces of the instantaneous pressure gradient magnitude. For clarity, the iso-surface threshold is varied for each aspect ratio to best highlight the leading shock structure. This normalization was necessary due to differences in shock strength and wall confinement, which alter the local pressure gradients and the sharpness of shock features across aspect ratios. For the circular duct ($AR = 1.0$, Fig.~\ref{fig.instant3D}a), the leading shock forms a nearly axisymmetric structure with a prominent normal shock stem at the center, surrounded by azimuthally distributed oblique shock segments. A normal shock train is expected for the current flow conditions and blockage (see Fig.2 in \citet{hunt2018shock}). The irregularities at the periphery of the oblique shock segments result from their interaction with the turbulent boundary layer~\cite{hunt2018shock,carroll1990characteristics}. As the aspect ratio increases, the leading shock becomes progressively anisotropic, with the shock front elongating along the duct’s major axis. The extent of the normal shock stem is significantly reduced and nearly vanishes at $AR=3.0$ (Fig.~\ref{fig.instant3D}c). 

\begin{figure*}[!ht]
	\centering
   \includegraphics[trim={0cm 0.25cm 0cm 0},clip,width=.95\textwidth]{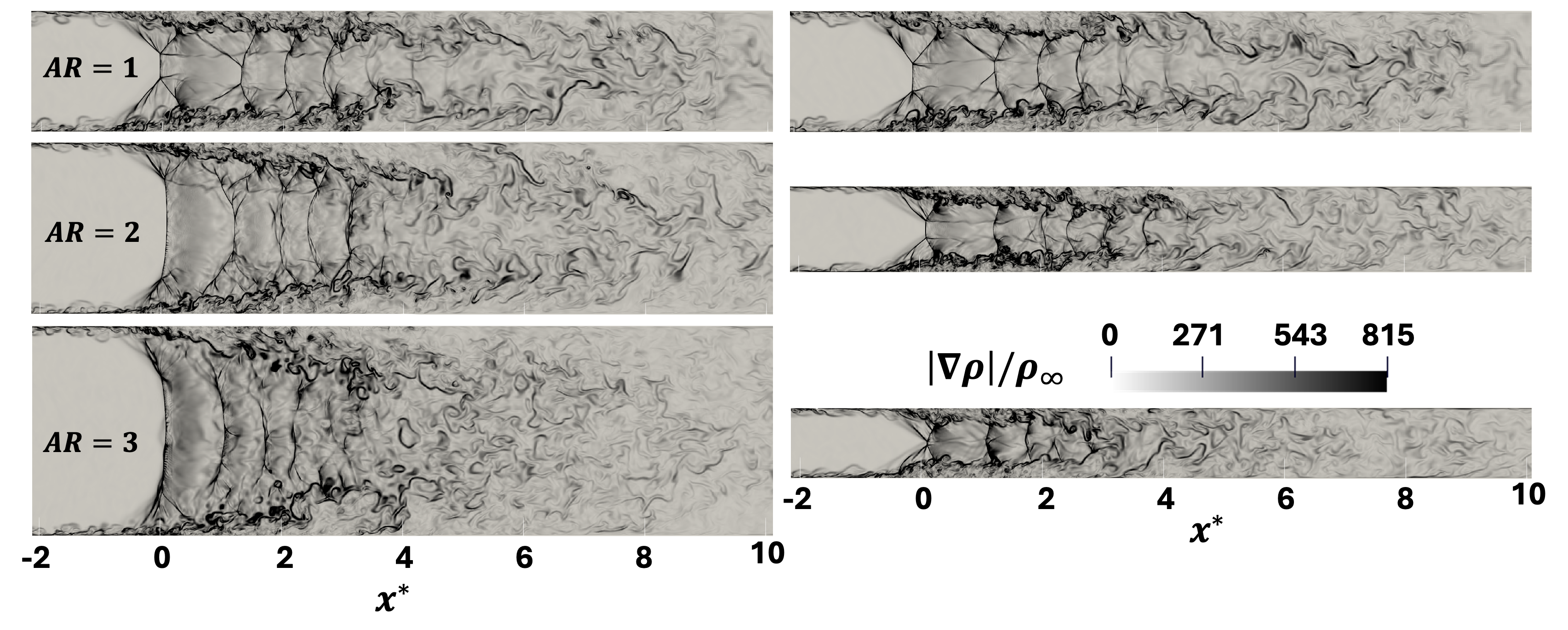}  
\caption{shock trains visualized via non-dimensional instantaneous density gradient magnitude $|\nabla \rho|/\rho_\infty$ plotted on the center z-plane (on left) and on the center y-plane (right) for (from top to bottom)  $AR=1.0$,  $AR=2.0$ and $AR=3.0$.}
	\label{fig.instant2D}
\end{figure*}

\begin{figure*}[!ht]
	\centering
   \includegraphics[trim={0cm 0.25cm 0cm 0},clip,width=.95\textwidth]{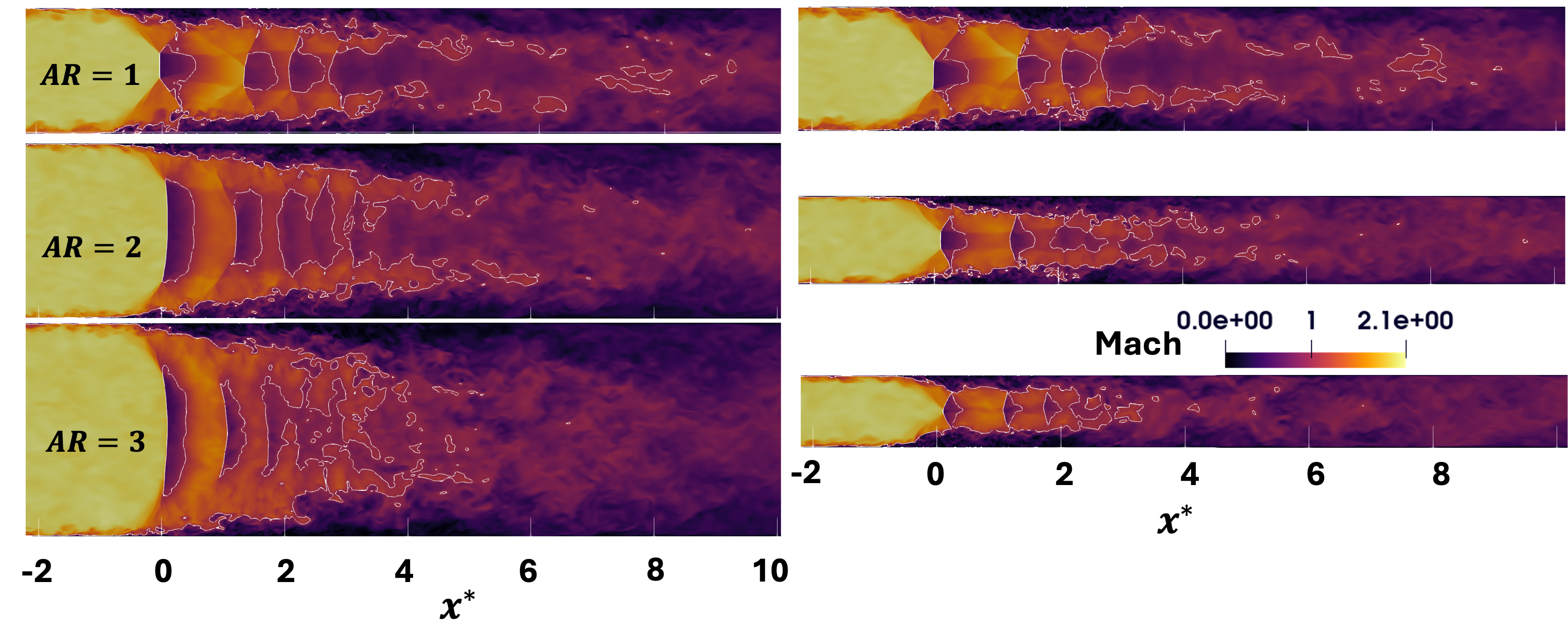}  
\caption{Instantaneous Mach number plotted on the center z-plane (left) and on the center y-plane (right) for (from top to bottom)   $AR=1.0$,  $AR=2.0$ and $AR=3.0$. Contour lines of $M=1$ are shown in white.}
	\label{fig.instant2DMach}
\end{figure*}

This evolution is seen more clearly in Fig.~\ref{fig.instant2D}, which shows instantaneous density gradient contours on the center $y$- and $z$-planes. These planes capture the development of the shock train along the duct’s major and minor axes, respectively. The shock train appears as a series of discrete shock cells with repeating patterns. Following separation induced by the primary shock, the boundary layer grows rapidly and contibuting more towards the pressure rise which leads to a gradual reduction in the strength and definition of subsequent shocks. This effect is evident for $AR = 1.0$ and $AR = 2.0$, where multiple successive shock cells are clearly identifiable. As seen earlier in Fig.~\ref{fig.instant3D}, with increasing $AR$, the shock cells become elongated along the duct's major axis and compressed along the minor axis. For $AR = 3.0$, confinement along the minor axis appears to be sufficient to suppress the formation of a prominent normal shock stem. Figure~\ref{fig.instant2DMach} complements these observations by presenting instantaneous Mach number contours on the same center planes. The $M = 1$ contour marks the transition between supersonic and subsonic flow. As $AR$ increases, the subsonic core region behind the shock train expands more rapidly and the overall extent of the shock train shortens. The separation between consecutive shock cells decreases as $AR$ increases. This behavior contrasts with the effect of increasing pressure ratio across the pseudo shock, which reduces the number of shock cells, and with increasing boundary-layer thickness, which increases the spacing between shock cells~\cite{singh2024shock,gillespie2022shock}.

\begin{figure*}[!ht]
	\centering
  \includegraphics[trim={0.0cm 0.2cm 0cm 0cm},clip,width=.95\textwidth]{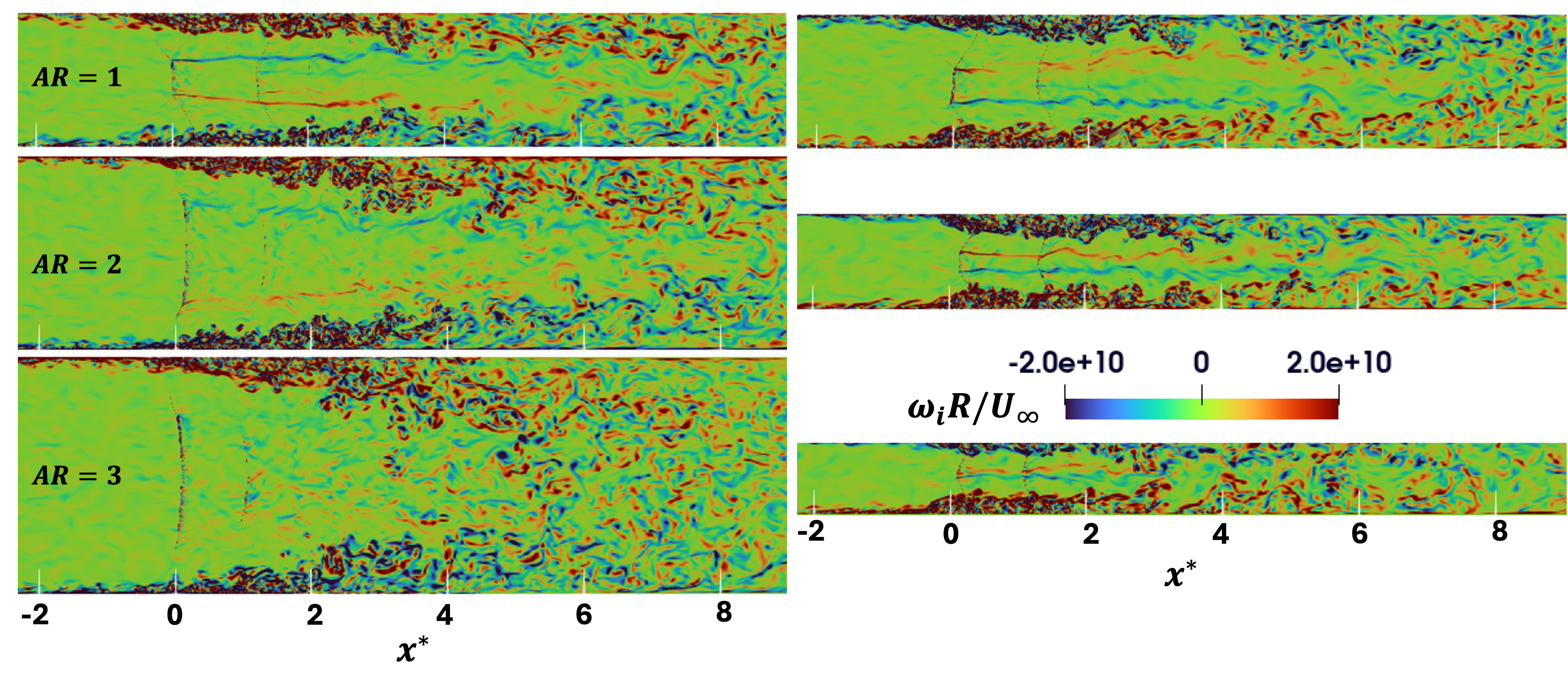}
\caption{Normalized instantaneousness vorticity component $\omega_z$ plotted on the center z-plane (left) and $\omega_y$ plotted on the center y-plane (right) for (from top to bottom)   $AR=1.0$,  $AR=2.0$ and $AR=3.0$.}
	\label{fig.vort2D}
\end{figure*}

To understand the effect of geometry on shock-induced vorticity, the spatial distribution of the instantaneous vorticity $\bm{\omega}=\bm{\nabla} \times \bm{u}$ normalized by $R/U_\infty$ is shown in Fig.~\ref{fig.vort2D} on $z-$ and $y-$ center planes. Only the out-of-plane components i.e. $\omega_z$ for $z-$ plane and $\omega_y$ for $y$-plane, are shown. A common feature across all configurations is the formation of concentrated vorticity along a slip line where the normal shock stem and oblique shock front meet. This slip line arises from a discontinuity in the post-shock flow direction, giving rise to a shear layer that subsequently roll up into vortical structures. With increasing $AR$, both the spatial extent and intensity of this slip line weaken. At $AR = 3.0$, it nearly vanishes on the $z$-plane (Fig.~\ref{fig.vort2D}a,c,e), and similarly contracts on the $y$-plane due to the shrinking footprint of the central normal shock. Nonetheless, residual slip-line vorticity remains visible in the $y$-plane even at high aspect ratios. Although the slip-line vorticity is consistently present in all three cases, the dominant contribution to vorticity arises from near-wall regions where shock-induced boundary layer thickening.  Higher $AR$ increases geometric confinement along the minor axis, which enhances wall-driven turbulence and allows vortices to extend deeper into the duct core. This effect is particularly evident at $AR = 3.0$, where large-scale vortical structures can be seen clearly on both planes close to the axis downstream of the shock train (Fig.~\ref{fig.vort2D}e,f).
These vortical structures lead to additional flow deceleration downstream of the shock train implying that the pressure rise in high-$AR$ configurations is increasingly driven by subsonic mixing region rather than by discrete shocks. This is discussed in more details in the following section using time-averaged quantities.

\subsection{Mean flow structure and turbulent kinetic energy}

\begin{figure}[!ht]
    \centering
    \begin{tabular}{c}
    \makebox[0pt]{\raisebox{28ex}{\hspace{-2.8ex}{(a)}}} \includegraphics[width=0.8\linewidth]{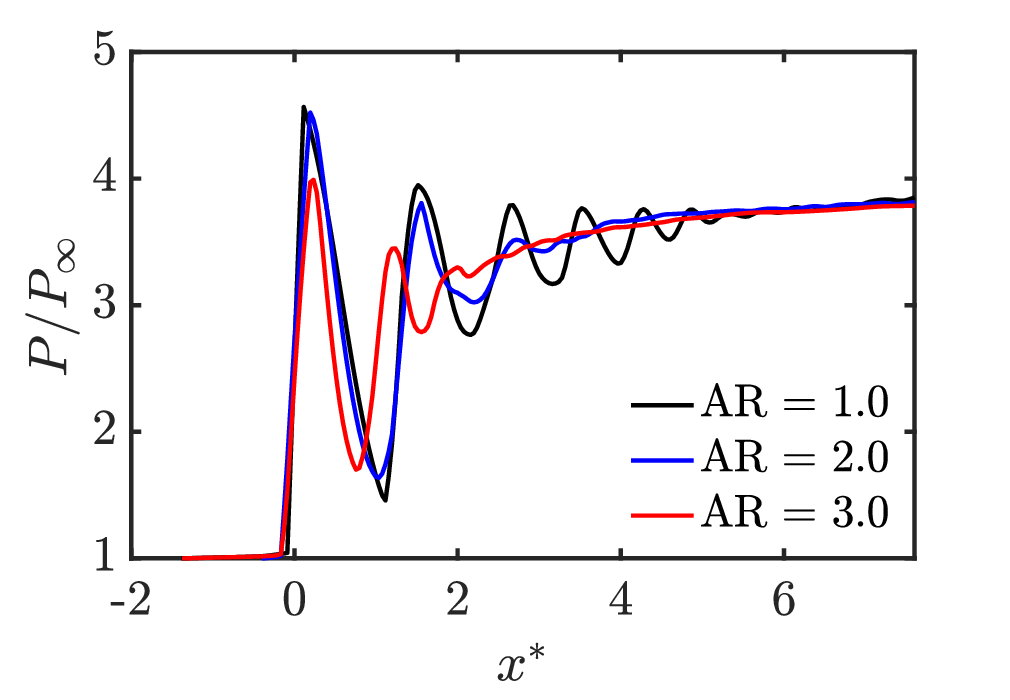} \\ 
   \makebox[0pt]{\raisebox{28ex}{\hspace{-2.8ex}{(b)}}} \includegraphics[width=0.8\linewidth]{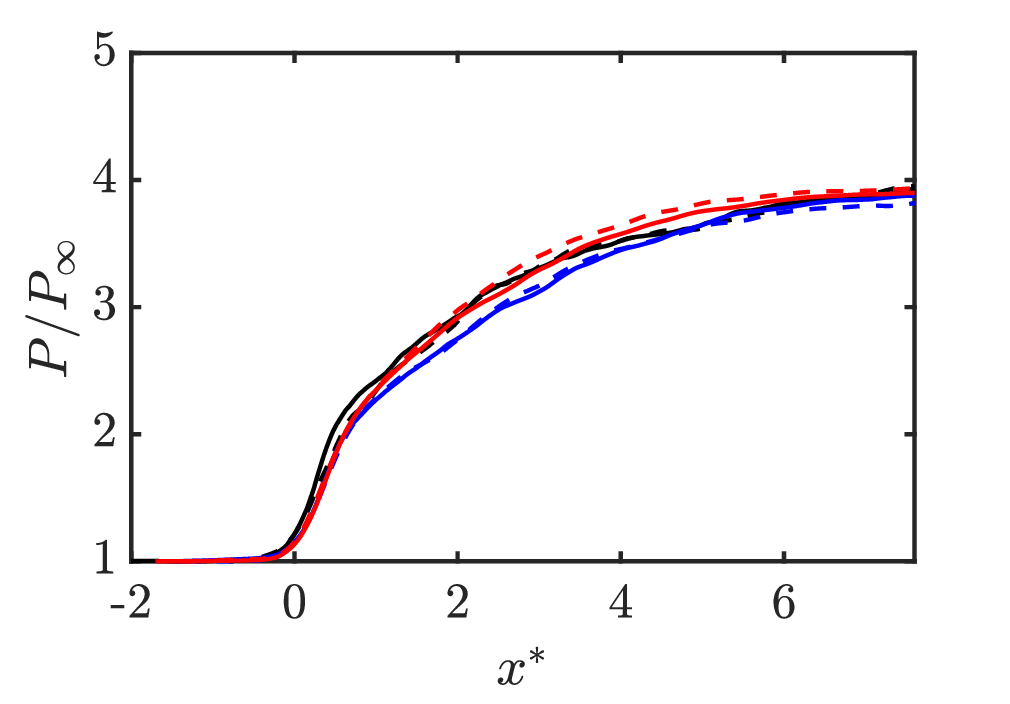} 
       \end{tabular}
        \caption{Comparison of the time-averaged pressure rise (a) at the axis and (b) at the wall for different $AR$. In (b) solid lines show the pressure rise along the major axis while dashed lines show pressure rise along the minor axis. }
    \label{fig.meanP}
\end{figure}

To assess the influence of duct geometry on the mean structure of the shock train and the associated pressure recovery, the time-averaged static pressure profiles along the duct centerline and wall are analyzed in Fig.~\ref{fig.meanP}. Along the centerline, where the flow undergoes a sequence of supersonic-to-subsonic transitions, the pressure profile exhibits distinct jumps corresponding to the individual shocks that constitute the shock train. In contrast, the wall pressure increases smoothly along the streamwise direction, reflecting the subsonic nature of the near-wall flow. As the aspect ratio ($AR$) increases, the number of discrete shock cells decreases, indicating a transition to more distributed compression driven primarily by turbulent mixing in the subsonic core. Traditionally, the term \emph{shock train} refers to the streamwise region where a sequence of shocks is visually identifiable, while the downstream portion characterized by continued pressure rise in a nearly subsonic regime due to turbulent entrainment and shear is referred to as the \emph{mixing region}. Together, these define the broader \emph{pseudo shock} structure~\cite{matsuo1999shock,hunt2018shock}. In the present simulations, the total static pressure rise is nearly identical across all $AR$ as imposed in the simulations (Fig.~\ref{fig.meanP}). Therefore, a decrease in the number of  shocks in the shock train with $AR$ indicates a redistribution of pressure recovery mechanisms, with a greater fraction of the total pressure rise occurring in the mixing region at higher $AR$.

\begin{figure}[!ht]
    \includegraphics[width=0.8\linewidth]{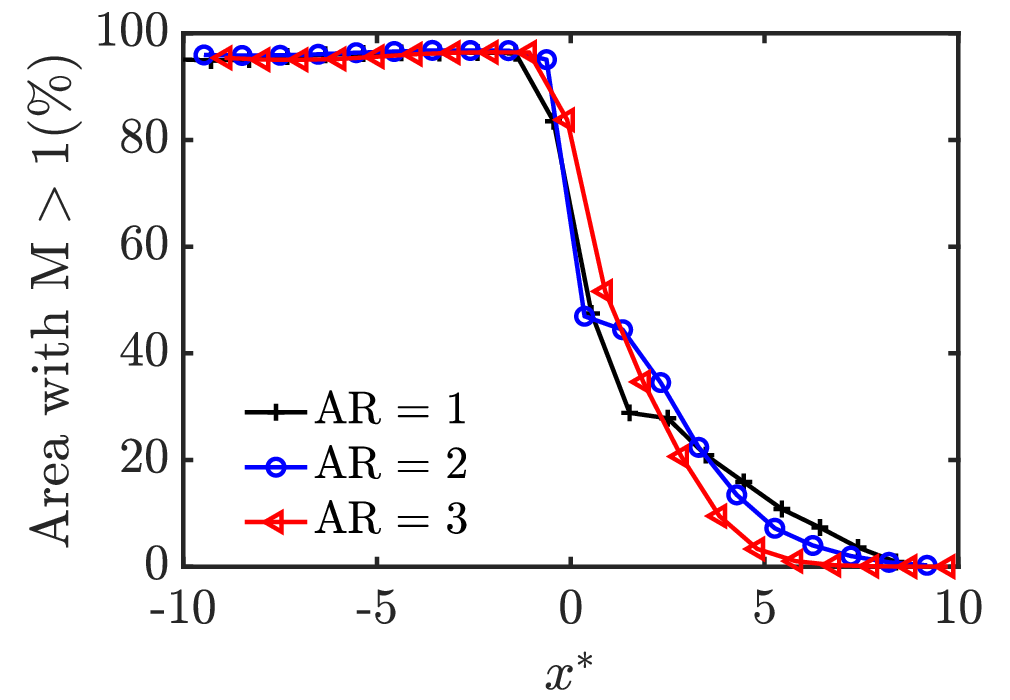} \\
         \caption{Comparison of (a) the fraction of cross-sectional area with $M>1$ relative to the total cross-sectional area xtand (b) area and time-averaged stagnation pressure ratio  computed at different x-locations for different $AR$.}
    \label{fig.integral}
\end{figure}

To quantify the spatial extent of supersonic flow, data was extracted on multiple $x$-planes during the simulation and the time-averaged area with $M>1$ was computed. The resulting variation of the supersonic cross-sectional area along the flowpath is shown in Fig.~\ref{fig.integral}. For $AR=3.0$, the supersonic core collapses completely by $x^*\approx6$, while for $AR=1.0$, supersonic flow persists until $x^*\approx9$. Despite these differences, the mean pressure at both the centerline and the wall converges across all $AR$ by $x^*\approx6$ (Fig.~\ref{fig.meanP}). This trend implies that while the total pseudo shock length remains relatively insensitive to $AR$, the contribution of discrete shock cells to the overall pressure rise decreases with increasing $AR$, suggesting that the remaining supersonic regions near the axis have limited influence on the integral pressure rise. 

\begin{figure}[!ht]
    \centering
           \includegraphics[trim={0cm 0.2cm 0cm 0cm}, clip, width=0.49\textwidth]{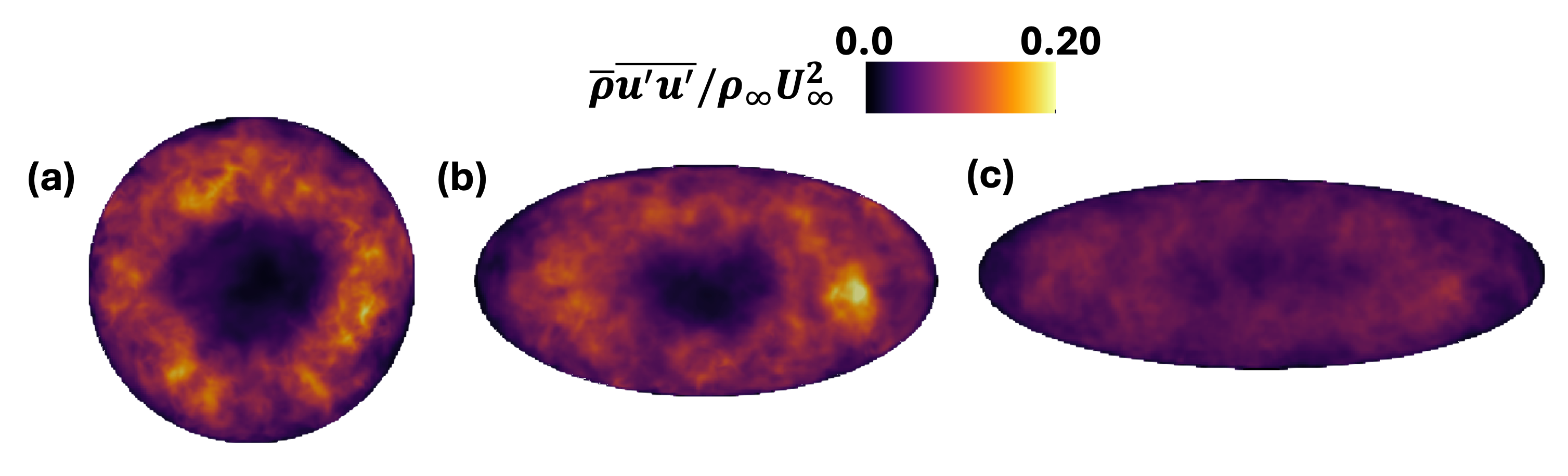}  
        \caption{Turbulent kinetic energy normalized by the freestream kinetic energy plotted at $x^*=5$ for (a) $AR = 1.0$, (b) $AR = 2.0$, and (c) $AR = 3.0$.}
    \label{fig.tke}
\end{figure}

\begin{figure}[!ht]
    \centering
 \includegraphics[width=0.8\linewidth]{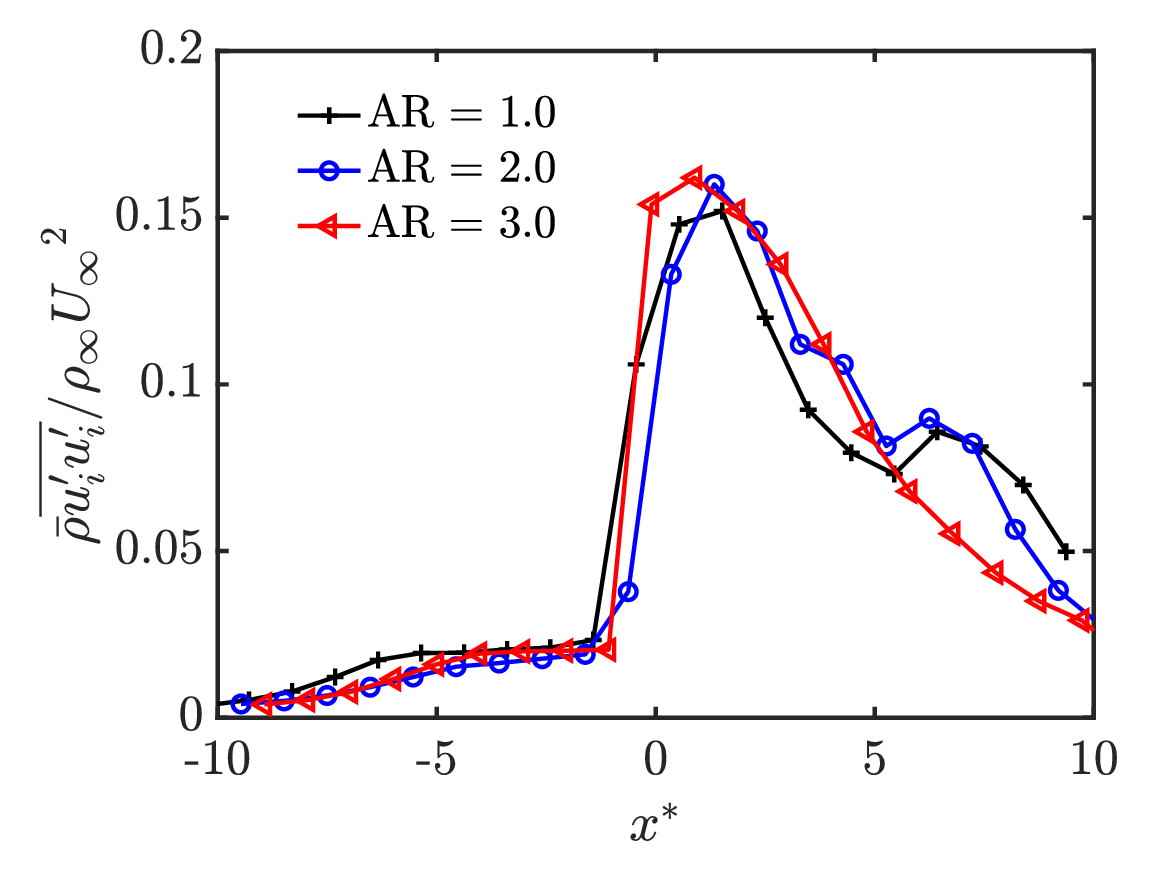} 
          \caption{Area averaged turbulent kinetic energy normalized by the freestream kinetic energy plotted for different $AR$. }
    \label{fig.TKE_int}
\end{figure}

To understand the effect of vortical structures in the core observed in Fig.~\ref{fig.instant2D} on overall turbulence levels, the turbulent kinetic energy (TKE) is analyzed at $x^* = 5$ in Fig.~\ref{fig.tke} where TKE is normalized by the freestream kinetic energy, while Fig.~\ref{fig.TKE_int} shows the corresponding area-averaged TKE extracted on several cross-sections along the flowpath. For $AR = 1.0$ and $AR = 2.0$, high TKE concentrations appear in an annular region in the middle of the flow area between wall and the axis. This is due to the interaction between the boundary layer and oblique shocks within the shock train. These interactions do not fully dissipate by $x^* = 5$. In contrast, for $AR = 3.0$, such localized high-TKE zones are largely absent, reflecting a more gradual pressure rise and weaker shock-boundary layer interactions as expected from more curved oblique shock front and lesser number of shock cells in the shock train. Figure~\ref{fig.TKE_int} further reveals that the onset of the shock train coincides with a pronounced spike in TKE reaching approximately 15\% of the freestream kinetic energy primarily due to rapid boundary layer thickening induced by the adverse pressure gradient caused by the leading shock and shock oscillations. A secondary rise in TKE around $x^* \approx 7$ for $AR = 1.0$ and $AR = 2.0$ is associated with the interaction between boundary layers from opposite walls, however, this spike quickly dissipates downstream. 

The stagnation pressure loss across the pseudo shock is governed by both shock-induced compression and viscous dissipation. To assess the effect of the geometry on this loss, the area-averaged stagnation pressure is computed at two streamwise planes: one located upstream of the leading shock at $x^*=-10$ and the other sufficiently downstream where the influence of $AR$ becomes negligible, $x^*=10$. The resulting stagnation pressure ratio $P_{0_2}/P_{0_1}$ where $P_{0_2}$ is calculated at $x^*=10$ and $P_{0_1}$ is calculated at $x^*=-10$, characterizes the total loss across the pseudo shock system. Interestingly, despite significant differences in the internal and three-dimensional shock structure, the relative contribution of the mixing region and TKE distribution, the stagnation pressure loss is nearly identical for all three $AR$ values ($P_{0,2}/P_{0,1}\approx 0.683$). This finding suggests that, for a fixed freestream Mach number and equivalent blockage due to boundary layer, the total pressure loss and pseudo shock length remain largely insensitive to the duct aspect ratio.

\section{Summary and Conclusions}

This study investigates the effect of duct aspect ratio ($AR$) on shock train structure and pressure recovery in elliptical ducts using AMR-EB framework. The central objective is to determine whether boundary-layer-induced blockage is the primary parameter governing geometric confinement, and to assess how the cross-sectional anisotropy offered by the elliptical geometry modifies the shock train behavior. Numerical simulations are performed for $AR = 1.0$ to $AR = 3.0$ at a freestream Mach number of 2.1, with a constant pressure ratio imposed across the shock train via a conical plug. The aspect ratio is varied while preserving the duct cross-sectional area, thereby ensuring that the incoming mass flow remains unchanged. A fixed boundary-layer thickness corresponding to $5\%$ of the radius of the circular duct ($AR = 1.0$) is imposed. This configuration results in an effective geometric blockage of approximately $9.8\%$ for $AR = 1.0$ and $11.2\%$ for $AR = 3.0$. Holding the area constant while increasing $AR$ naturally increases the duct perimeter by roughly $16\%$, which modifies the hydraulic diameter (see Table~\ref{tab.cellCount}). Simulations use nearly $600$ million finite-volume cells at a finest resolution of $48\,\mu\text{m}$ and are conducted on fifteen H100 GPUs at their finest resolution.

Results show that the shock train structure is strongly dependent on $AR$.  The leading shock exhibits a pronounced near-axisymmetric normal shock stem for $AR=1.0$, which weakens progressively and nearly disappears at $AR = 3.0$. The confinement along the minor axis in $AR=3.0$ produces sufficient blockage for the near-wall vortical structures to penetrate closer to the duct centerline than in the circular case. This redistribution of the flow enables additional flow deceleration and flow transitions to fully subsonic state earlier in $AR=3.0$ compared to $AR=1.0$. The number of shock cells decreases from approximately five at $AR = 1.0$ to two at $AR = 3.0$. Because the overall pressure ratio across the shock train is fixed, a direct implication is that a larger portion of the total pressure rise in high-$AR$ ducts occurs through mixing rather than through discrete shocks. The turbulent kinetic energy (TKE) downstream of the shock train also becomes more laterally uniform with increasing $AR$, although the cross-sectionally integrated TKE varies only slightly. Remarkably, despite these significant changes in shock train morphology, the wall-pressure distributions and stagnation-pressure losses across the pseudo shock which includes the mixing region, remain nearly identical across all $AR$. These findings indicate that while the local morphology of the shock train is sensitive to $AR$, integral response metrics such as wall-pressure rise, total stagnation-pressure loss, and cross-sectional TKE remain largely insensitive or only weakly sensitive to $AR$ when the key flow parameters - boundary-layer blockage, upstream Mach, and the overall pressure ratio across the pseudo shock are held constant. The observation that increasing aspect ratio substantially modifies the maximum local blockage along the minor axis and the hydraulic diameter, while leaving pressure recovery largely unchanged, suggests that neither maximum local blockage nor hydraulic diameter provides a reliable scaling parameter for shock train behavior in elliptical ducts in contrast to rectangular ducts~\cite{geerts2016shock,cox2001effect}. The current results are also in contrast with results obtained in rectangular ducts, where shock train length was reported to increase with $AR$~\cite{cox2001effect}. However, in those studies, $AR$ was modified by changing a single duct dimension, which simultaneously altered the inflow mass rate and boundary-layer blockage, making it difficult to isolate the effect of aspect ratio alone.

Finally, this work demonstrates the capability of the EB method with AMR to capture shock train dynamics in complex geometries. The current EB implementation requires uniform mesh spacing, which complicates achieving high near-wall resolution. In DNS of boundary layer flows, $y^+<1$ is warranted but the resolution in other directions can be significantly higher. Incorporating nonuniform meshing within the EB framework would enable more efficient simulations and allow quantitative comparison of the computational benefits of the EB+AMR approach. In current simulations for all $AR$, the shock train was observed to drift upstream at approximately $2\%$ of the core flow velocity, consistent with prior high-resolution studies~\cite{gillespie2023numerical,fievet2017numerical,qin2025coherent}. Whether the shock train ultimately reaches a stationary equilibrium remains an open question. Unlike rectangular ducts, which introduces corner-induced separation and strong secondary motions, elliptical ducts lack such geometric effects. A direct comparison of elliptical and rectangular geometries may therefore provide deeper insight into how cross-sectional shape influences shock train morphology and overall pressure rise. 

\begin{acknowledgments}
This work was supported by the US Office of Naval Research under Grant N00014-21-1-2475 with Dr. Eric Marineau as Program Manager and the US  Air Force Office of Scientific Research under grant number FA9550-23-1-0067 with Dr. Fariba Fahroo as Program Manager. IT support provided by the Lighthouse Advanced Research Computing team at the University of Michigan is greatly acknowledged.
\end{acknowledgments}

\section*{Data Availability Statement}
The data that support the findings of this study are available from the corresponding author upon reasonable request.

\nocite{*}
\bibliography{apssamp}

\end{document}